\documentclass[11pt, onecolumn]{IEEEtran}

\usepackage{graphicx} 
\usepackage{amsmath} 
\usepackage{amssymb}  
\usepackage{algorithm}
\usepackage{algorithmic}
\usepackage{cite}

\newtheorem{theorem}{Theorem}
\newtheorem{lemma}{Lemma}
\newtheorem{defn}{Definition}
\newtheorem{corollary}{Corollary}
\newtheorem{remark}{Remark}

\newcommand{\reals}{\mathbb{R}}

\newcommand{\Expect}{\mathbb{E}}
\newcommand{\expect}[1]{\mathbb{E}\left[#1\right]}
\newcommand{\Prob}{\mathbb{P}}
\newcommand{\prob}[1]{\mathbb{P}\left[#1\right]}

\newcommand{\pth}[1]{\left( #1 \right)}
\newcommand{\qth}[1]{\left[ #1 \right]}
\newcommand{\sth}[1]{\left\{ #1 \right\}}

\newcommand{\indc}[1]{{\mathbf{1}\left\{{#1}\right\}}}

\newcommand{\calD}{{\mathcal{D}}}
\newcommand{\calE}{{\mathcal{E}}}

\newcommand{\calI}{{\mathcal{I}}}

\newcommand{\calL}{{\mathcal{L}}}

\newcommand{\calV}{{\mathcal{V}}}

\newcommand{\calX}{{\mathcal{X}}}
\newcommand{\calY}{{\mathcal{Y}}}

\newcommand{\bfR}{{\mathbf{R}}}

\newcommand{\bfX}{{\mathbf{X}}}
\newcommand{\bfY}{{\mathbf{Y}}}

\newcommand{\bfx}{{\mathbf{x}}}
\newcommand{\bfy}{{\mathbf{y}}}

\renewcommand{\iff}{\Leftrightarrow}

\newcommand{\floor}[1]{{\left\lfloor {#1} \right \rfloor}}

\begin{document}
\title{Finite-Sample Analysis of Image Registration} 

\author{Ravi Kiran Raman and Lav R.\ Varshney}

\maketitle

\begin{abstract}
We study the problem of image registration in the finite-resolution regime and characterize the error probability of algorithms as a function of properties of the transformation and the image capture noise. Specifically, we define a channel-aware Feinstein decoder to obtain upper bounds on the minimum achievable error probability under finite resolution. We specifically focus on the higher-order terms and use Berry-Esseen type CLTs to obtain a stronger characterization of the achievability condition for the problem. Then, we derive a strong type-counting result to characterize the performance of the MMI decoder in terms of the maximum likelihood decoder, in a simplified setting of the problem. We then describe how this analysis, when related to the results from the channel-aware context provide stronger characterization of the finite-sample performance of universal image registration.
\end{abstract}

\section{Introduction}

Image registration is the task of geometrically aligning two or more images of the same scene taken at different points in time, from different viewpoints, or by different imaging devices. For instance, given different scans of an anatomy, such as MRI, CT, and X-ray, one might be interested in a comparative study across the scans. Since the imaging modalities and image capture instances are different, these images however, even though a copy of the same scene, differ fundamentally in the image content, and orientation. Thus one requires an algorithm to correctly align these image copies to one common orientation. This task is common in several domains such as medical imaging \cite{DuTJL2006}, cryo-electron microscopy \cite{ZhaoS2014}, and remote sensing \cite{ChenAV2003}. 

The problem has been extensively studied and a wide range of pixel- and feature-based methods have been devised to address the problem. We mention a few popular approaches. Among pixel-based methods, information and distance-based methods are of particular interest \cite{ViolaW1997, AlhichriK2001, PluimMV2003, ZitovaF2003, ChanCYNW2003, ChenDV2019}. Fast Fourier transforms have also been used as features to perform scale-invariant image registration efficiently \cite{ReddyC1996, AverbuchK2002, Guizar-SicairosTF2008, TzimiropoulosAZS2010}. More recently, neural networks have been employed in both unsupervised and weakly supervised settings to design multimodal and non-rigid image registation algorithms by extracting the latent features of the image copies \cite{SimonovskyGMNK2016, deVosBVSI2017, SokootiVBLI2017, EppenhofLMVP2018}. Such algorithms have been shown to perform well on practical contexts, especially extensively in medical imaging. However results on the fundamental theoretical characterization of the problem have been limited \cite{ZolleiW2009, XuCV2009, TagareR2015, AguerrebereDBS2016}. 

Theoretical analyses of information processing systems not only help us understand the optimality properties of existing practical algorithms, but also inspire novel algorithms. Conventional information-theoretic investigations of communications study the channel capacity (mean) and error exponent (large deviations) of noisy channels. Similarly, our recent explorations in \cite{RamanV2018c} focused on asymptotic optimality of universal image registration algorithms in terms of the error exponent. In particular, the analysis establishes that the channel-agnostic max mutual information (MMI) method achieves the same error exponent as achieved by the Bayes optimal maximum likelihood detector. Further, this type-counting based study of the max mutual information method also highlighted the robustness of information functionals, inspiring the universally asymptotically optimal multi-image registration algorithm called the max multiinformation method \cite{RamanV2018c}.

Whereas this analysis shows the universality and asymptotic optimality of information-based registration algorithms, it fails to characterize the performance of these algorithms in the non-asymptotic regime, i.e., when the image resolution is limited. Further, the task of image registration is fundamentally reliant on the image capture noise, i.e., the channel that defines the relation between the images. Naturally, the noisier the copies, the harder the alignment. Thus, it is important to study the effects of the image capture noise on algorithm performance. The optimal error exponents of image registration were quantified and analyzed in \cite{Raman2019_thesis}. We draw inspiration from more recent information theoretic studies to perform a stronger analysis for finite-resolution images.

Going beyond asymptotic analyses, information-theoretic studies have been dedicated to understanding second-order terms such as the bivariate information-theoretic quantity called channel dispersion, using Berry-Esseen type central limit theorems \cite{Berry1941, Esseen1942}, to allow better understanding of non-asymptotic performance limits of compression and communication \cite{Weiss1960, Strassen1962, Polyanskiy2010, KostinaV2013}. In lossless and lossy compression of known source models, tight results on the higher-order terms have been obtained \cite{Strassen1962, Hayashi2008, KostinaV2012, Kostina2013, KontoyiannisV2014, Tan2014}.

Extensive characterization of the non-asymptotic fundamental limits of communication have also been established \cite{Kontoyiannis1997, Polyanskiy2010, PolyanskiyPV2010, TanM2014}. Second-order methods with sharp non-asymptotics have also been derived in designing constrained encoding and decoding for communication \cite{TanM2014, Moulin2017}. Of particular interest is the set of results on non-asymptotic fundamental limits for constant composition coding \cite{Hayashi2009, Moulin2012, KostinaV2015}. A variety of finite-sample performance analyses have also been performed for statistical inference \cite{IngberWK2012, TomamichelT2013, TomamichelH2013, HanN2017}. In particular, the conditional error probabilities for Neyman-Pearson (NP) binary hypothesis test in the finite-sample setting have been studied in detail using the Berry-Esseen theorem \cite{Strassen1962, PolyanskiyPV2010}. Strong large deviations analyses informed by the Cramer-Esseen theorem have also been used in deriving strong converses in the finite-sample regime \cite{Moulin2012, HuangM2014}.

Theoretical analyses, especially in the finite-sample context, have however been concentrated on the channel and source aware contexts mostly. In practice we require efficient universal methods for compression, communication, and statistical inference as the statistical information for the context are mostly unavailable. Universal information theoretic studies have mostly studied the asymptotic setting \cite{LapidothN1998, Csiszar1998, Misra2015, RamanV2018c, RamanV2017a}. Some recent results have established second- and third-order error terms for universal lossless compression under fixed blocklength coding \cite{Kontoyiannis2000, KosutS2017}. Such studies in the context of statistical inference have however been limited. In this paper, we aim to provide an outline of an approach to finite-sample characterization of universal methods by obtaining strong bounds with respect to the optimal channel-aware decoders, and subsequently leveraging their second order analyses.

In this paper we aim to characterize the performance of image-registration algorithms in the finite-sample regime, characterized as a function of the channel characteristics. specifically, we build achievability arguments using a likelihood-based registration algorithm for images with finite resolution. In particular, we build on the second-order studies mentioned above to identify the sufficient condition on the image resolution for achieving an alignment error probability of $\epsilon$ as a function of the moments of the information spectrum of the channel.

Then, in order to characterize the performance of universal methods, we study the simplified, one-dimensional version of the image registration problem, the universal delay estimation problem \cite{SteinZM1996}. In this context, we study the MMI decoder using a strong type counting argument to benchmark its performance, as a function of the sample size $n$, with respect to that of the Bayes optimal, channel-aware, maximum likelihood decoder. In particular, the characterization highlights the sub-optimality of the MMI method in relation to ML in terms of the properties of the transformations to be detected. This characterization helps establish the finite-sample performance of universal methods, given strong characterizations of the performance of likelihood-based decoders. And given the relationship between the delay estimation problem and the image registration problem, we briefly highlight how these results could be extended to study universal image registration.

The paper is organized as follows. First in Sec.~\ref{sec:system_model} we define the image, channel, and noise model that we study, and also define the loss functions and formal performance metrics. In Sec.~\ref{sec:channel_aware_second} we study an achievability condition of finite-sample performance of a likelihood-based registration algorithm, establishing sufficient conditions on the sample complexity. Finally we study a strong type counting argument for universal delay estimation in Sec.~\ref{sec:univ_delay_est} to provide a roadmap to characterizing the finite sample performance of universal methods in terms of those of channel-aware optimal methods, before concluding in Sec.~\ref{sec:conclusion} with some possible future research directions.

\section{System Model} \label{sec:system_model}

\begin{figure}[t]
	\centering
	\includegraphics[scale=0.3]{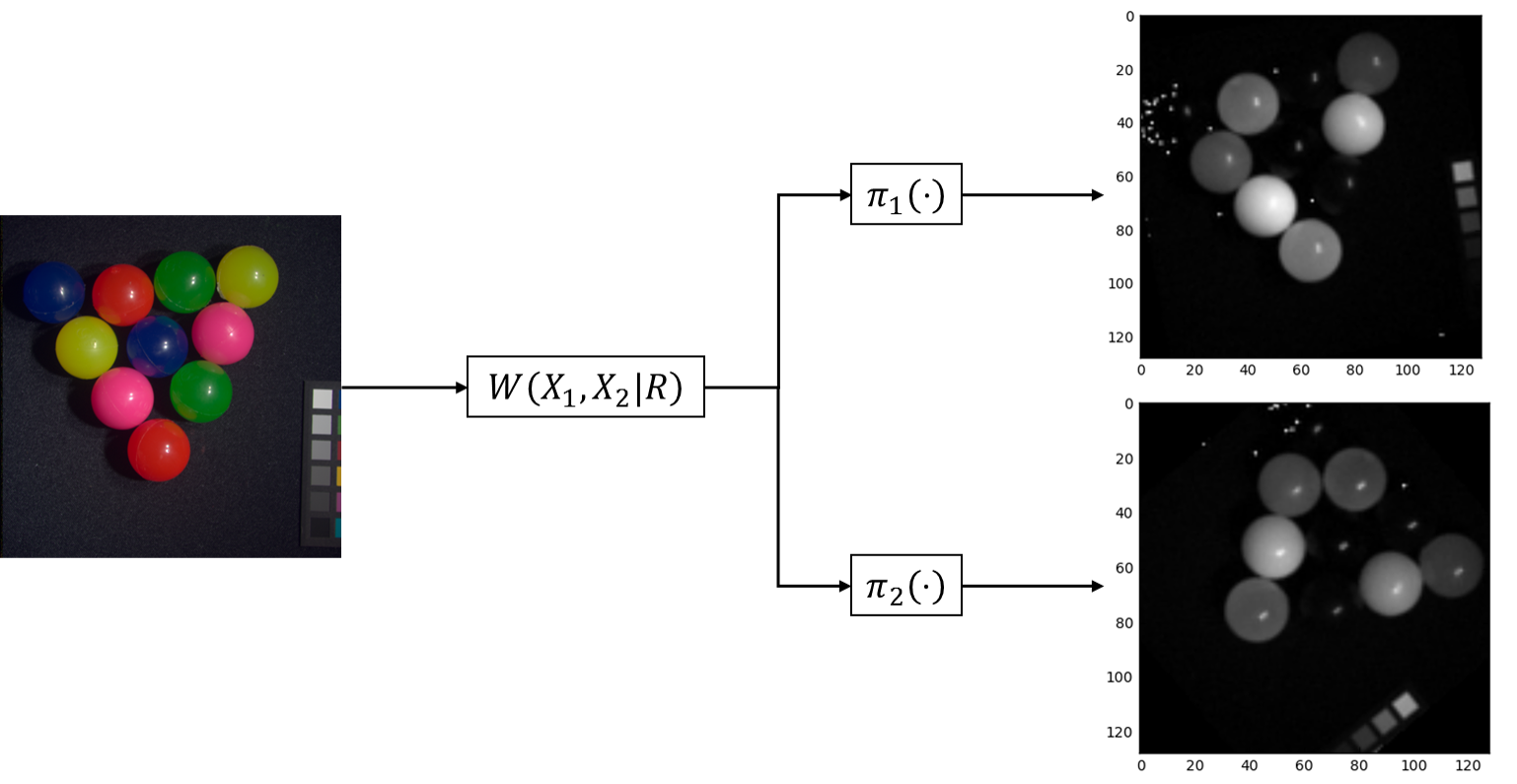}
 	\caption{Model of the image registration problem: Pixels of the underlying scene are jointly corrupted by DMC $W$ and images are transformed by rigid body transformations $\pi_1,\pi_2$.}
  	\label{fig:image_model}
\end{figure}

We consider a simple image model, wherein each image is a noisy version of a collection of $n$ pixels drawn independently and identically from an unknown prior defined on a finite set of pixel values $[r] = \{1,\dots,r\}$, as depicted in Fig.~\ref{fig:image_model}.

Let the scene captured by an image be an $n$-dimensional random vector, $\bfR \sim P_R^{\otimes n}$. Consider two images of the scene, each of which is a noisy depiction (channel output) of the scene (source), i.e., outputs of a discrete memoryless channel (DMC) whose input is the scene:
\begin{align}
&\prob{\tilde{\bfX}, \tilde{\bfY} \middle\vert \bfR} = \prod_{i = 1}^{n} W\pth{\tilde{X}_i, \tilde{Y}_i \middle\vert R_i}. \label{eqn:image_channel_model}
\end{align}
That is, images are jointly corrupted by a DMC, and the pixels of the images are independent of each other. Without loss of generality, we assume $\tilde{\bfX} \in [r]^n$. 

In this work we consider an i.i.d.\ source to model the images and exclude inter-pixel correlations. Whereas the correctness of the forthcoming registration algorithms holds for ergodic sources, they are much harder to analyze rigorously. Further, we also limit our study to pixel-based registration for ease of description. Such algorithms have also been adapted to work on higher-order features of images \cite{NeemuchwalaHZC2006}. Thus, even if pixels are not i.i.d.,\ the assumption remains reasonable at the feature level, and the results of the analysis are broadly retained.

Since images are modeled with i.i.d.\ pixels, distinguishing alignments is similar to independence testing. The Type-1 and Type-2 error exponents of independence testing are the mutual and lautum information \cite{PalomarV2008} respectively. Thus, we presume without loss of generality that $L(X;Y) \leq L_{\text{max}} < \infty$.

These corrupted images are subject to rigid-body transformations of rotation and translation on the discrete $\mathbb{Z}^2$-lattice. Conventional methods consider rotation by an angle of $\theta \in [0,2\pi)$ followed by a translation of $[t_x,t_y]' \in \reals^2$. Then, the discrete equivalent of these transformations for a pixel at location $[x,y]'$ is
\[
\pi([x,y]') = \calD \pth{\qth{\begin{matrix}
\cos \theta & -\sin \theta \\
\sin \theta & \cos \theta
\end{matrix}} \qth{\begin{matrix}
x \\ y
\end{matrix}} + \qth{\begin{matrix}
t_x \\ t_y
\end{matrix}}},
\]
where $\calD:\reals^2 \to \mathbb{Z}^2$ is the rounding function. In general such mappings are not bijective, owing to the rounding operation. In practice however, they are considered to be invertible using a standard backwards mapping: $\pi^{-1} = \calD\circ\pi^{-1}$ \cite{NgoKPT2015}.

In this work, instead of this continuous transformation and resampling model for transforming images, we adopt a permutation model as it generalizes the set of all rigid-body transformations and is also more conducive to study algorithms that work at pixel-scale. Since images are vectors of length $n$, we represent the transformations by permutations of $[n]$. Let $\pi_j \sim \text{Unif}(\Pi)$ be the transformation of image $j$. Then, the final image is $\bfX^{(j)}_{i} = \tilde{\bfX}^{(j)}_{\pi_j(i)}$, for all $i \in [n]$. Image $\bfX$ transformed by $\pi$ is depicted interchangeably as $\pi(\bfX) = \bfX_{\pi}$. Let the set of all discrete rigid body transformations be $\Pi$. The number of such transformations of an image with $n$ pixels on the $\mathbb{Z}^2$-lattice is $M=|\Pi| = o(n^5)$ \cite{NgoKPT2015}. We assume $\Pi$ is known.

We assume $\Pi$ forms a \emph{commutative algebra} over the composition operator $\circ$. More specifically,
\begin{itemize}
\item for $\pi_1,\pi_2 \in \Pi$, $\pi_1 \circ \pi_2 = \pi_2 \circ \pi_1 \in \Pi$;
\item there exists unique $\pi_0 \in \Pi$ s.t. $\pi_0(i) = i$, for all $i \in [n]$;
\item for any $\pi \in \Pi$, there exists a unique inverse $\pi^{-1} \in \Pi$, s.t. $\pi^{-1} \circ \pi = \pi \circ \pi^{-1} = \pi_0$.
\end{itemize}
Studies on image registration typically presume the application of a rotation followed by a translation, in order. Thus, the assumption of the permutations forming a commutative algebra does not remove anything from the problem.

\begin{defn} \label{defn:opt_regn}
The \emph{correct registration} of an image $\bfX$ transformed by $\pi \in \Pi$ is $\hat{\pi} = \pi^{-1}$.
\end{defn}

\begin{defn} \label{defn:permut_props}
Let us define some permutation-related terms.
\begin{itemize}
\item A \emph{permutation cycle} of $\pi \in \Pi$ is a subset $\{i_1,\dots,i_k\}$ of $[n]$, such that $\pi(i_{j}) = i_{j+1}$, for all $j < k$ and $\pi(i_k) = i_1$. Let the number of permutation cycles of $\pi$ be $\kappa_\pi$.
\item \emph{Identity block} of $\pi \in \Pi$ is the inclusion-wise maximal subset $\calI_{\pi}$ of $[n]$ such that $\pi(i) = i$, for all $i \in \calI_{\pi}$.
\item A permutation $\pi$ is a \emph{derangement} if $\kappa_{\pi} = 1$, $\calI_{\pi} = \emptyset$.
\end{itemize}
\end{defn}

We now introduce formal performance metrics.
\begin{defn}
A \emph{universal image registration algorithm} is a sequence of functions, $\Phi^{(n)}: [r]^2 \rightarrow \Pi^2$, designed in the absence of knowledge of $W$ and $P_R$. Here, $n$ corresponds to the number of pixels in each image.	
\end{defn}
We focus on the $0$-$1$ loss function to quantify performance. 
\begin{defn}
The \emph{error probability} of an algorithm $\Phi^{(n)}$ that outputs $\pth{\hat{\pi}_1,\hat{\pi}_2} \in \Pi^2$ is
\begin{align}
P_e(\Phi^{(n)}) &= \prob{\cup_{i \in [2]} \{\hat{\pi}_i \neq \pi_i^{-1}\}}. \label{eqn:regn_error_eq}
\end{align} 
\end{defn}

Finally, the Bayes optimal detector is the maximum likelihood detector, given by
\begin{equation}\label{eqn:ML_regn}
\hat{\pi}_{\text{ML}} = \arg\max_{\pi\in\Pi} \prod_{i=1}^n W\pth{Y_{\pi(i)} \vert X_i}.
\end{equation}
The max mutual information (MMI) method for image registration is a universal method that has been shown to be asyptotically optimal \cite{RamanV2018c}. The detector is defined as
\begin{equation} \label{eqn:MMI_regn}
\hat{\pi}_{\text{MMI}} = \arg\max_{\pi\in\Pi} \hat{I}(X;Y_{\pi}),
\end{equation}
where $\hat{I}(X;Y)$ is the empirical mutual information.

\section{Channel-Aware Image Registration} \label{sec:channel_aware_second}

To establish the fundamental performance limits for image registration in the finite-sample context we now consider channel-aware two-image registration. In particular, we define and study an image registration algorithm based on the information density. 

\subsection{Moments of Information Density}

For this chapter we will characterize the performance in terms of the moments of information density, and we introduce the moments here for reference. 

\begin{defn} \label{defn:info_density}
Given $X, Y\sim p_{X,Y}(\cdot)$, with corresponding marginals $p_X,p_Y$, the \emph{information density}, also called \emph{information spectrum}, is defined as
\[
\imath(x;y) = \log \frac{p_{X,Y}(x,y)}{p_X(x)p_Y(y)}.
\]
\end{defn}
The moments of the information density are as follows:
\begin{enumerate}
\item Mutual information, $I(X;Y) = \expect{\imath(X;Y)}$.
\item Dispersion, $V(X;Y) = \expect{(\imath(X;Y) - I(X;Y))^2}$.
\item Third absolute moment, $T(X;Y) = \expect{ | \imath(X;Y) - I(X;Y) |^3}$.
\end{enumerate}
Several properties of these moments have been studied. In particular, we note that the information density, mutual information, dispersion, and third absolute moments are continuous over the probability simplex. Further,  the dispersion and the third absolute moment are bounded above \cite{PolyanskiyPV2010}.

\subsection{The Feinstein Decoder}

The two-image registration problem has been well-studied and a variety of registration algorithms have been defined, including the MMI decoder. Since we consider the Hamming loss, the Bayes optimal algorithm is the maximum likelihood estimate.  Whereas the ML decoder is Bayes optimal, to assist with the analysis, we consider the Feinstein version of a likelihood ratio test to perform the registration \cite{Feinstein1954}. To define this decoder, let us presume that the possible transformations are ordered as $\Pi = \sth{\pi_{i}: i\in [M]}$. Then the transformation is estimated as
\begin{equation}\label{eqn:Feinstein_regn}
\hat{\pi}_{\text{F}} = \pi_{i^*}, \text{ where } i^* = \min\sth{i \in [M]: \imath(\bfX; \bfY_{\pi_i}) \geq \delta}.
\end{equation}
The Feinstein decoder is a version of the likelihood ratio test, as proved by the following lemma.
\begin{lemma}
For any image pair, $\bfX, \bfY$, and transformation $\pi$, 
\[
\imath(\bfX, \bfY_{\pi}) = L_{\pi}(\bfX,\bfY) + C(\bfX,\bfY),
\]
where $L_{\pi}(\cdot)$ is the log likelihood ratio given transformation $\pi \in \Pi$, and $C(\bfX,\bfY) = - \log \pth{p(\bfX)p(\bfY)}$, a function independent of $\pi$.
\end{lemma}
\begin{IEEEproof}
The result follows directly from the definition of the information density and the memorylessness of the channel and source.
\end{IEEEproof}

We note here that the Feinstein decoder is also closely related to the MMI decoder owing to the following result.
\begin{lemma} \label{lemma:density_mi_relation}
For any discrete source $P$, if $(X_i,Y_i) \sim P$ for $i\in [n]$, and $\hat{P}$ is the corresponding empirical distribution, then
\begin{equation}
    \hat{I}(X;Y) - \frac{1}{n}\imath(\bfX;\bfY) = D(\hat{P}\|P) - D(\hat{P}_X \| P_X) - D(\hat{P}_Y \| P_Y), \label{eqn:KL_Diff_density}
\end{equation}
where $P_X, P_Y, \hat{P}_X, \hat{P}_Y$ are the marginal distributions.
\end{lemma}
\begin{IEEEproof}
We skip the subscripts where evident for ease of description. The result follows by regrouping the terms in the definition of empirical mutual information as follows.
\begin{align}
    \hat{I}(X;Y) &= D\pth{\hat{P} || \hat{P}_X \hat{P}_Y } \notag \\
    &= \Expect_{\hat{P}} \qth{\log \pth{\frac{\hat{P}(X,Y)}{\hat{P}(X) \hat{P}(Y)} }} \notag \\
    &= \frac{1}{n} \sum_{i = 1}^n \log \pth{\frac{\hat{P}(x_i,y_i)}{\hat{P}_X(x_i) \hat{P}_Y(y_i)} } \notag \\
    &= \Expect_{\hat{P}} \qth{\log \pth{\frac{\hat{P}(X;Y)}{P(X;Y)}} - \log \pth{\frac{\hat{P}(X)}{P(X)}} - \log \pth{\frac{\hat{P}(Y)}{P(Y)}}} +  \frac{1}{n} \sum_{i = 1}^n \log \pth{\frac{P(x_i,y_i)}{P(x_i) P(y_i)} } \notag \\
    &= D(\hat{P}\|P) - D(\hat{P}_X \| P_X) - D(\hat{P}_Y \| P_Y) + \imath (\bfX;\bfY). \notag
\end{align}
\end{IEEEproof}
Since the empirical distributions converge to the true distribution, and since the mutual information is a continuous function on the probability simplex, we know that both the average information density and empirical mutual information converge to the mutual information. More precisely, the function thresholded by the Feinstein decoder is approximately equal to the empirical mutual information for large $n$. This is useful for future studies as we could extend performance results on Feinstein-type decoders to the universal algorithms by deriving strong bounds on the KL divergence terms in \eqref{eqn:KL_Diff_density}.

The closeness of the Feinstein decoder to MMI also implies that one could implement the algorithm in practice without explicit knowledge of the channel, using appropriate estimates of the information \cite{BelghaziBROBCH2018, RahimzamaniAVK2018, GaoOV2018}. The practical difficulty with the Feinstein decoder however is the search algorithm. In its current form, it performs exhaustive search of the transformation space which is not feasible in practice. However, just as with MMI, we can develop gradient descent based heuristics to accelerate the search.

\subsection{Achievability Arguments}

We now derive upper bounds on the error probability of the Feinstein decoder for image registration in the channel-aware, finite-sample context. In particular we characterize the tradeoff between the sample size (image resolution) and channel properties (moments of information density) under which the decoder achieves an error probability of $\epsilon$.

First, the error probability of the Feinstein decoder can be decomposed into two main components as follows.
\begin{theorem} \label{thm:Feinstein_sum_prob_bound}
The error probability of the Feinstein decoder is bounded as
\begin{equation}\label{eqn:sum_prob_bound}
P_{e}(\Phi_{F}) \leq \Prob_{\pi_0}\qth{\imath(\bfX;\bfY) \leq \delta} + \frac{M-1}{2} \Prob_{\pi_0}\qth{\imath(\bfX;\bfY_{\pi'}) > \delta},
\end{equation}
where $\pi'$ is the transformation with the maximum number of fixed points.
\end{theorem}
\begin{IEEEproof}
First, we bound the conditional error probability, given the true transformation is $\pi^* = \pi_j$. In this case, the decoder declares the wrong transformation if $\imath(\bfX;\bfY_{\pi_j}) \leq \delta$ or there exists $i<j$ such that $\imath(\bfX;\bfY_{\pi_i}) > \delta$, i.e.,
\begin{align}
P_{e,j} &= \prob{\hat{\pi}_F \neq \pi_j \vert \pi^* = \pi_j} \notag \\
&= \prob{\sth{\imath(\bfX;\bfY_{\pi_j}) \leq \delta} \cup_{i<j} \sth{\imath(\bfX;\bfY_{\pi_i}) > \delta} \vert \pi^* = \pi_j} \notag \\
&\leq \Prob_{\pi_j}\qth{\imath(\bfX;\bfY_{\pi_j}) \leq \delta} + \sum_{i<j} \Prob_{\pi_j}\qth{\imath(\bfX;\bfY_{\pi_i}) > \delta} \label{eqn:union_bd_cases}\\
&= \Prob_{\pi_0}\qth{\imath(\bfX;\bfY) \leq \delta} + \sum_{i<j} \Prob_{\pi_j}\qth{\imath(\bfX;\bfY_{\pi_i}) > \delta} \label{eqn:true_hyp_dist}\\
&\leq \Prob_{\pi_0}\qth{\imath(\bfX;\bfY) \leq \delta} + \sum_{i<j} \Prob_{\pi_0}\qth{\imath(\bfX;\bfY_{\pi'}) > \delta} \label{eqn:worst_hyp_bound} \\
&= \Prob_{\pi_0}\qth{\imath(\bfX;\bfY) \leq \delta} + (j-1) \Prob_{\pi_0}\qth{\imath(\bfX;\bfY_{\pi'}) > \delta}, \label{eqn:sum_bound_j}
\end{align}
where \eqref{eqn:union_bd_cases} follows from the union bound, \eqref{eqn:true_hyp_dist} follows since the information density between the correctly transformed pairs conditioned on the true transformation is the same as that between the given image pairs under the null hypothesis. Finally, \eqref{eqn:worst_hyp_bound} is obtained by bounding the conditional probabilities by the transformation with the most fixed points with respect to the null hypothesis. This is since it has the least informative samples, and thus serves as a bound for the probability.

Finally, \eqref{eqn:sum_prob_bound} follows from \eqref{eqn:sum_bound_j} as
\begin{align*}
P_e(\Phi_F) &= \sum_{j \in [M]} \frac{1}{M} P_{e,j} \\
&\leq \Prob_{\pi_0}\qth{\imath(\bfX;\bfY) \leq \delta} + \frac{1}{M} \binom{M}{2} \Prob_{\pi_0}\qth{\imath(\bfX;\bfY_{\pi}) > \delta}.
\end{align*}
\end{IEEEproof}
Let us elaborate on the terms of this decomposition. In Thm.~\ref{thm:Feinstein_sum_prob_bound}, the first term, $\Prob_{\pi_0}\qth{\imath(\bfX;\bfY) \leq \delta}$ is the cumulative distribution function of the information density of $n$ i.i.d.\ samples $(X_i,Y_i) \sim P$, evaluated at $\delta$. That is, it is the probability that the information density in $n$ samples drawn from the joint distribution $P$ is smaller than the threshold $\delta$. Given the mutual information $I$ corresponding to the joint distribution $P$, we know that the normalized information density, i.e., $\tfrac{1}{n} \imath(\bfX,\bfY)$, concentrates around $I$. Thus, for a threshold $\delta$ that scales much faster than $nI$, this error term converges to $0$.

Next, the second term of the decomposition is given by $\Prob_{\pi_0}\qth{\imath(\bfX;\bfY_{\pi'}) > \delta}$. Here, $\pi'$ is specified to be the transformation with the maximum number of fixed points. This effectively implies that the transformation consists of the most number of dependent pairs $(X_i, Y_{\pi'(i)}) \sim P$. All other pairs $(X,Y)$ corresponding to non-fixed points of the transformation are independent of each other. Note however that these samples are still dependent on the corresponding pixel from the image copy. Hence the error term evaluates the tail probability of the information density in such samples that consist largely of independent pairs and so converges to $0$ with the sample size.

Next, we bound the two probabilities in \eqref{eqn:sum_prob_bound}.
\begin{lemma} \label{lemma:error_1_bound}
Given $n$ i.i.d. pairs $(\bfX,\bfY) \stackrel{i.i.d.}{\sim} p_{X,Y}$,
\begin{equation}\label{eqn:inf_density_Berry_bd}
\prob{\imath(\bfX;\bfY) \leq \delta} \leq Q(\tau) + \frac{B}{\sqrt{n}},
\end{equation}
where
\[
\tau = \frac{n I(X;Y) - \delta}{\sqrt{n V(X;Y)}}, \text{ and } B = \frac{6T(X;Y)}{V(X;Y)^{3/2}}.
\]
\end{lemma}
\begin{IEEEproof}
Let $Z_i = \imath(X_i;Y_i)$. Then, $\imath(\bfX;\bfY) = \sum_{i=1}^n Z_i$, and for any $i \in [n]$,
\begin{align*}
\mu_Z& = \expect{Z_i} = I(X;Y), \\
V_Z &= \text{var}(Z_i) = \expect{(Z_i - \mu_Z)^2} = V(X;Y), \\
T_Z &= \expect{|Z_i - \mu_Z|^3} = T(X;Y).
\end{align*}
Then, from the Berry-Esseen theorem, we have
\begin{align*}
\prob{\imath(\bfX;\bfY) \leq \delta} &= \prob{\sum_{i=1}^n Z_i \leq \delta} \\
&\leq Q\pth{\frac{n \mu_Z - \delta}{\sqrt{n V_Z}}} + \frac{6T_Z}{\sqrt{n} (V_Z)^{3/2}},
\end{align*}
and the result is obtained by substituting the values of the computed moments.
\end{IEEEproof}
This lemma directly characterizes the first term in \eqref{eqn:sum_prob_bound}.

On the other hand, to study the second term in \eqref{eqn:sum_prob_bound}, we first study the tail probabilities of the information densities corresponding to the fixed points, and that corresponding to the derangement separately. Sanov's theorem results in the following upper bound on the tail probability of the information density.
\begin{lemma} \label{lemma:Sanov_inf_density_tail}
Given, $n$ i.i.d. samples $(X_i,Y_i) \stackrel{i.i.d.}{\sim} p$, then for any constant $\lambda > 0$,
\begin{equation} \label{eqn:Sanov_info_ub}
\prob{\imath(\bfX;\bfY) \geq n \delta} \leq C_n \exp\pth{-n \qth{\lambda I - \log \expect{\exp\pth{\lambda \imath}}}},
\end{equation}
where $I = I(X;Y), \imath = \imath(X;Y), C_n = (n+1)^{|\calX||\calY|}$. The tightest upper bound from the inequality is obtained by using
\[
\lambda^* = \arg\max_{\lambda > 0} \lambda I(X;Y) - \log \expect{\exp \pth{\lambda \imath(X;Y)}}.
\]
\end{lemma}
\begin{IEEEproof}
The proof uses Sanov's theorem. For simplicity, let $Z_i = \imath(X_i;Y_i)$. Then, from Sanov's theorem, we have
\[
\prob{\imath(\bfX;\bfY) \geq n \delta} \leq (n+1)^{|\calX||\calY|} \exp\pth{-n D(q^* \| p)},
\]
where 
\[
q^* = \arg\max_{q: \Expect_q\qth{\imath(X;Y)} \geq \delta} D(q \| p).
\]

Using Lagrange multiplier $\lambda > 0$, consider the Lagrangian
\begin{align*}
\calL(q) &= D(q\|p) - \lambda \Expect_q\qth{\imath(X;Y)} \\
&= \sum_{(x,y) \in \calX\times\calY} q(x,y) \qth{\log \frac{q(x,y)}{p(x,y)} - \lambda \log \frac{p(x,y)}{p(x)p(y)}}.
\end{align*}

To maximize the Lagrangian, we set the partial derivatives to $0$, and have
\begin{align*}
& \frac{\partial}{\partial q(x,y)} = 0 \iff \log \frac{q(x,y)}{p(x,y)} - \lambda \frac{p(x,y)}{p(x)p(y)} + 1 = 0 \\
&\iff  q(x,y) = \frac{1}{Z} p(x,y) \qth{\frac{p(x,y)}{p(x)p(y)}}^{\lambda},
\end{align*}
where 
\[
Z = \expect{\pth{\frac{p(x,y)}{p(x)p(y)}}^{\lambda}},
\]
is the normalization constant. Since the optimization is over a convex objective with linear inequality constraints, from KKT conditions, it is evident that
\[
\delta = \Expect_{q^*} \qth{\imath(X;Y)} = \frac{1}{Z} \Expect_p\qth{\pth{\frac{p(X,Y)}{p(X)p(Y)}}^{\lambda} \log \pth{\frac{p(X,Y)}{p(X)p(Y)}}}.
\]

Thus, we can compute the maximum KL divergence as
\begin{align*}
D(q^*\|p) &= \Expect_{q^*} \qth{\log \frac{q^*(x,y)}{p(x,y)}} \\
&= \frac{1}{Z} \Expect_{p}\qth{\pth{\frac{p(X,Y)}{p(X)p(Y)}}^{\lambda} \log \pth{\frac{1}{Z} \pth{\frac{p(X,Y)}{p(X)p(Y)}}^{\lambda}}} \\
&= \lambda \delta - \log Z \\
&= \lambda \delta - \log \expect{\exp\pth{\lambda \imath(X;Y)}}.
\end{align*}

Now, let $\tilde{X} =\exp\pth{\lambda \imath(X;Y)}$. Then, we have
\[
\lambda \delta = \frac{\expect{\tilde{X} \log \tilde{X}}}{\expect{\tilde{X}}}.
\]
Since the function $f(x) = x\log x$ is convex, using Jensen's inequality, we have
\[
\lambda \delta \geq \frac{\expect{\tilde{X}} \log \expect{\tilde{X}}}{\expect{\tilde{X}}} \geq \log \pth{\exp \pth{\lambda \expect{\imath(X;Y)}}} = \lambda I(X;Y).
\]
Thus, for any $\lambda>0$, $\delta \geq I(X;Y)$ and so the result follows.
\end{IEEEproof}

Finally, let us consider the information density generated by pairs of pixels in a derangement by using the Berry-Esseen theorem.
\begin{lemma} \label{lemma:derangement_density}
Consider $n$ i.i.d. pairs $(X_i,Y_i) \stackrel{i.i.d.}{\sim} p$ and let $\pi$ be a derangement of $[n]$. That is, for all $i \in [n]$, $\pi(i) \neq i$. Then, for $\imath(\bfX,\bfY_{\pi}) = \sum_{i=1}^{n} \imath(X_i;Y_{\pi(i)})$, the tail probability is bounded as
\begin{align}
    \prob{\imath(\bfX,\bfY_{\pi}) \geq \delta} \leq 6\sqrt{3} \pth{\frac{\log 2}{\sqrt{2\pi}} + 2B(X;Y)} \frac{1}{\sqrt{n V(X;Y)}} \exp\pth{-\frac{\delta}{3}}.
\end{align}

\end{lemma}
\begin{IEEEproof}
Note that for any $i$, $(X_i,Y_{\pi(i)}) \sim p_X p_Y$. However, note that the samples themselves are dependent as we are considering permutations of the sequence. So we first split the samples into sets of independent pairs. First construct a graph based on the permutation $\pi$ on the set of vertices $V = [n]$ with edges $(i,\pi(i))$, for all $i \in [n]$. Since the permutation is a derangement, the resulting graph is composed of a set of disjoint cycles, each of length at least two. Thus the vertices are $3$-colorable. By uniformly distributing the three colors among the nodes, divide the set as $[n] = \calV_1 \cup \calV_2 \cup \calV_3$, according to the colors of the corresponding nodes in the graph. It is easy to see that there exists a coloring such that $|\calV_i| \geq \floor{\frac{n}{3}}$ for all $i\in [3]$. For simplicity, we assume that $n$ is a multiple of $3$ and that $|\calV_i| = n/3$. The results generalize trivially.

Since $\calV_i$ includes nodes of the same color, for any $j,k \in \calV_i$, it is evident that $\pi(j),\pi(k) \notin \calV_i$. Consequently, $(X_j,Y_{\pi(j)})$ and $(X_{k}, Y_{\pi(k)})$ are independent. More generally, the pairs corresponding to the indices in any $\calV_i$ are mutually independent and $(X_j,Y_{\pi(j)}) \stackrel{i.i.d.}{\sim} p_X p_Y$, for any $j \in \calV_i$.

We first note that
\begin{align}
\prob{\imath(\bfX,\bfY_{\pi}) \geq \delta} &= p_{X,Y}^{\otimes n} \qth{\sum_{i\in[n]} \imath(X_i, Y_{\pi(i)}) \geq \delta} \notag \\
&\quad\leq \sum_{i \in [3]} p_{X,Y}^{\otimes n} \qth{\sum_{j \in \calV_i]} \imath(X_j, Y_{\pi(j)}) \geq \frac{\delta}{3}} \label{eqn:union_bd_derange} \\
&\quad= 3 p_Xp_Y^{\otimes \tfrac{n}{3}} \qth{\sum_{i \in [n/3]} \imath(X_i;Y_i) \geq \frac{\delta}{3} } \label{eqn:equiv_dist}\\
&\quad= 3\Expect_{p_{X,Y}^{\otimes \tfrac{n}{3} }} \qth{ \frac{p_X(\bfX)p_Y(\bfY)}{p_{X,Y}(\bfX,\bfY)} \indc{\sum_{i \in [n/3]} \imath(X_i;Y_i) \geq \frac{\delta}{3} }} \label{eqn:prob_change} \\
&\quad= 3 \Expect_{p_{X,Y}^{\otimes \tfrac{n}{3}} } \qth{\exp\pth{-\sum_{i \in [n/3]} \imath(X_i;Y_i)}  \indc{\sum_{i \in [n/3]} \imath(X_i;Y_i) \geq \frac{\delta}{3} }}, \label{eqn:info_density_defn}
\end{align}
where \eqref{eqn:union_bd_derange} follows from the union bound, \eqref{eqn:equiv_dist} follows from the fact that the probabilities are the same across the three color sets and that samples from each set are sampled independently according to the product distribution. Then, we change the distribution over which the expectation is computed to the joint distribution by appropriately scaling the indicator random variable, and finally, \eqref{eqn:info_density_defn} follows from the definition of the information density. Thus we have upper bounded the tail probability of the information density of the derangement by an indicator-weighted moment of the information density of samples drawn according to the joint distribution $p_{X,Y}$.

From \cite[Lemma 47]{PolyanskiyPV2010}, given $n$ independent random variables $Z_1,\dots,Z_n$, and if $V_Z = \sum_{j \in [n]} \text{Var} (Z_j) \neq 0$, and $T_Z = \sum_{j \in [n]} \expect{|Z_j - \expect{Z_j}|^3} < \infty$, then for any $\delta$, 
\[
\expect{\exp\pth{-\sum_{j \in [n]} Z_j} \indc{\sum_{j \in [n]} Z_j > \delta}} \leq 2 \pth{\frac{\log 2}{\sqrt{2\pi}} + \frac{12 T_Z}{V_Z}} \frac{1}{\sqrt{V_Z}} \exp\pth{-\delta}.
\]
Thus, if $Z_i = \imath(X_i;Y_i)$, the term in the upper bound in \eqref{eqn:info_density_defn} is bounded as
\begin{align*}
\Expect_{p_{X,Y}^{\otimes \tfrac{n}{3}} } \qth{\exp\pth{-\sum_{i \in [\tfrac{n}{3}]} \imath(X_i;Y_i)}  \indc{\sum_{i \in [\tfrac{n}{3}]} \imath(X_i;Y_i) \geq \frac{\delta}{3} }} \leq \pth{\frac{\log 2}{\sqrt{2\pi}} + 2B(X;Y)} \frac{2\sqrt{3}}{\sqrt{nV(X;Y)}} \exp\pth{-\frac{\delta}{3}},
\end{align*}
which proves the lemma.
\end{IEEEproof}

\begin{remark}
Note that the constant in Lem.~\ref{lemma:derangement_density} can be reduced to $4\sqrt{2}$ from $6\sqrt{3}$ if we knew the permutation cycles generated by $\pi$ were all of even length. This is because even length cycles can be vertex colored using two colors, resulting in two sets of size $n/2$.
\end{remark}

We can now use the results of Lems.~\ref{lemma:error_1_bound}, \ref{lemma:Sanov_inf_density_tail}, and \ref{lemma:derangement_density} to obtain the achievability criterion.
\begin{theorem} \label{thm:Feinstein_ub}
Let $M = 2c n^{\alpha} + 1$ and $\delta \geq 3\alpha \log n + \gamma_n n I(X;Y) + 3 \log c$. Then, the probability of error achieved by the Feinstein decoder is upper bounded as
\begin{align} \label{eqn:Feinstein_ub}
P_e(\Phi_F) &\leq Q\pth{\frac{nI - \delta}{\sqrt{nV}}} + \frac{B}{\sqrt{n}} + \frac{6\sqrt{3}}{\sqrt{(1-\gamma_n)n}} \pth{\frac{\log 2}{\sqrt{2\pi}} + 2B} + \frac{M-1}{2} (\gamma_nn+1)^{|\calX||\calY|}\exp\pth{-\gamma_n n D^*},
\end{align}
where 
\begin{align*}
I &= I(X;Y) = \expect{\imath(X;Y)}, \\
V &= V(X;Y) = \text{Var}\pth{\imath(X;Y)}, \\
T &= T(X;Y) = \expect{|\imath(X;Y) - I|^3}, \\
B &= B(X;Y) = \frac{6T}{V}, \\
D^* &= \lambda^* I(X;Y) - \log \expect{\exp\pth{\lambda^* \imath(X;Y)}}.
\end{align*}
\end{theorem}
\begin{IEEEproof}
From Thm.~\ref{thm:Feinstein_sum_prob_bound}, we know that 
\[
P_{e}(\Phi_{F}) \leq \Prob_{\pi_0}\qth{\imath(\bfX;\bfY) \leq \delta} + \frac{M-1}{2} \Prob_{\pi}\qth{\imath(\bfX;\bfY_{\pi}) > \delta}.
\]
From Lem.~\ref{lemma:error_1_bound}, we have
\[
\Prob_{\pi_0}\qth{\imath(\bfX;\bfY) \leq \delta} \leq Q\pth{\frac{nI - \delta}{\sqrt{nV}}} + \frac{B}{\sqrt{n}}.
\]

Next, from the union bound, we have
\begin{align}
\Prob_{\pi}\qth{\imath(\bfX;\bfY_{\pi}) > \delta} &= \Prob_{\pi}\qth{\sum_{j \in [n]} \imath(X_j; Y_{\pi(j)}) > \delta} \notag \\
&= \Prob_{\pi}\qth{\sum_{j \in \calI_{\pi}} \imath(X_j; Y_j) + \sum_{j \in \calI_{\pi}^{c}} \imath(X_j; Y_{\pi(j)}) > \delta} \notag \\
&\leq \Prob_{\pi}\qth{\sum_{j \in \calI_{\pi}} \imath(X_j; Y_j) \geq \delta_1} + \Prob_{\pi}\qth{\sum_{j \in \calI_{\pi}^{c}} \imath(X_j; Y_j) > \delta_2}, \label{eqn:error_cases_sum}
\end{align}
where $\delta_1 + \delta_2 = \delta$, and $\delta_1 \geq \gamma_n n I(X;Y)$, $\delta_2 \geq 3\alpha \log n + 3 \log c$.

From Lem.~\ref{lemma:Sanov_inf_density_tail}, for $\delta_1 \geq \gamma_n n I(X;Y)$, 
\[
\Prob_{\pi}\qth{\sum_{j \in \calI_{\pi}} \imath(X_j; Y_j) \geq \delta_1} \leq (\gamma_nn+1)^{|\calX||\calY|}\exp\pth{-\gamma_n n D^*}.
\]
Next, from Lem.~\ref{lemma:derangement_density}, for $\delta_2 \geq 3\alpha \log n + 3 \log c$,
\[
\Prob_{\pi}\qth{\sum_{j \in \calI_{\pi}^{c}} \imath(X_j; Y_j) > \delta_2} \leq \frac{2}{M-1} \frac{6\sqrt{3}}{\sqrt{(1-\gamma_n)n}} \pth{\frac{\log 2}{\sqrt{2\pi}} + 2B}.
\]
Substituting the probabilities, the result follows.
\end{IEEEproof}

Using the achievability criterion of Thm~\ref{thm:Feinstein_ub} we can obtain the sufficient condition on the image resolution to achieve an image registration error of $\epsilon$ as follows.
\begin{corollary} \label{cor:achievability}
If  $\tfrac{\log(1+\gamma_n n)}{\gamma_n n} \leq \tfrac{D^*}{2|\calX| |\calY|}$, and
\begin{equation} \label{eqn:achievability_condn}
(1-\gamma_n) n I(X;Y) \geq \sqrt{n V(X;Y)} Q^{-1}(\epsilon) + 3 \alpha \log n + \Delta,
\end{equation}
where $\Delta$ is a constant independent of the sample size $n$, but is dependent on the dispersion of the channel $V(X;Y)$, then, there exists a threshold $\delta$ such that the Feinstein decoder achieves an average error probability less than $\epsilon$.
\end{corollary}
\begin{IEEEproof}
First, from Thm.~\ref{thm:Feinstein_ub} we know that if $\delta_1 \geq \gamma_n n I(X;Y)$, $\delta_2 \geq 3\alpha \log n + 3 \log c$, then the probability of error of the Feinstein decoder is upper bounded as in \eqref{eqn:Feinstein_ub}. Thus, 
\begin{equation} \label{eqn:delta_lb}
\delta \geq \gamma_n n I(X;Y) + 3\alpha \log n + 3 \log c.
\end{equation}

Next, we note that if 
\begin{equation} \label{eqn:gamman_condn}
\gamma_n n \pth{D^* - O\pth{\tfrac{\log (\gamma_n n)}{\gamma_n n}}} \geq \pth{\alpha + \tfrac{1}{2}} \log n,
\end{equation}
then 
\[
\frac{M-1}{2} (\gamma_nn+1)^{|\calX||\calY|}\exp\pth{-\gamma_n n D^*} \leq \frac{c}{\sqrt{n}}.
\]

Since an achievability criterion for a larger $\gamma_n$ is also one for a smaller worst-case number of fixed points, it suffices to consider the case where $\gamma_n n \geq \frac{2\pth{\alpha + \tfrac{1}{2}}}{D^*} \log n$. Since $\tfrac{\log(1+\gamma_n n)}{\gamma_n n} \leq \tfrac{D^*}{2|\calX| |\calY|}$, \eqref{eqn:gamman_condn} is satisfied. Thus,
\[
P_e(\Phi_F) \leq Q\pth{\frac{nI - \delta}{\sqrt{nV}}} + \frac{B}{\sqrt{n}} + \frac{6\sqrt{3}}{\sqrt{(1-\gamma_n)n}} \pth{\frac{\log 2}{\sqrt{2\pi}} + 2B} + \frac{c}{\sqrt{n}}.
\]
Thus, $P_e(\Phi_F) \leq \epsilon$, if 
\begin{align}
\delta &\leq nI - \sqrt{nV} Q^{-1}\pth{\epsilon - \qth{\frac{B}{\sqrt{n}} + \frac{6\sqrt{3}}{\sqrt{(1-\gamma_n)n}} \pth{\frac{\log 2}{\sqrt{2\pi}} + 2B} + \frac{c}{\sqrt{n}}}} \notag \\
&= nI(X;Y) - \sqrt{n V(X;Y)} Q^{-1}(\epsilon) + \Delta, \label{eqn:delta_ub}
\end{align}
where 
\begin{align}
\eta\pth{\epsilon - \qth{\frac{B}{\sqrt{n}} + \frac{6\sqrt{3}}{\sqrt{(1-\gamma_n)n}} \pth{\frac{\log 2}{\sqrt{2\pi}} + 2B} + \frac{c}{\sqrt{n}}}} \leq \frac{\Delta}{B\sqrt{V} + \frac{6\sqrt{3}}{\sqrt{1-\gamma_n}}\pth{\frac{\log 2}{\sqrt{2\pi}} + 2B} + c} \leq \eta(\epsilon), \notag
\end{align}
where $\eta(\cdot)$ is the derivative of the inverse Q-function. Here \eqref{eqn:delta_ub} follows from the differentiability of the $Q^{-1}$ function.

Thus, from \eqref{eqn:delta_lb} and \eqref{eqn:delta_ub}, it is evident that an optimal threshold $\delta$ can be chosen provided \eqref{eqn:achievability_condn} holds.
\end{IEEEproof}

\begin{figure}[t]
	\centering
	\includegraphics[scale=0.7]{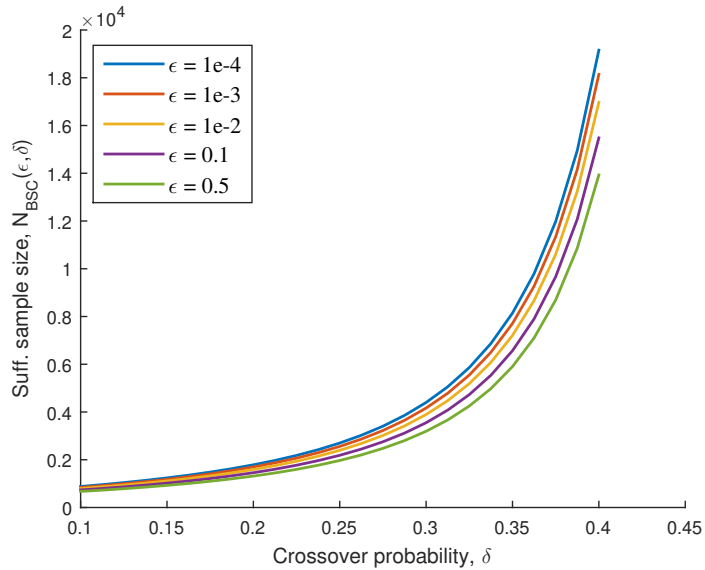}
 	\caption[Sample Complexity of BSC]{\small Sufficient sample size $N(\epsilon,\delta)$ for BSC($\delta$) and error probability $\epsilon$. The sufficient sample size increases with a decrease in target error probability, and with increasing crossover probability.}
  	\label{fig:second_order_bsc}
\end{figure}

Corollary~\ref{cor:achievability} characterizes the tradeoff between the channel properties, in terms of the moments of information density, and the sample size. To get a clearer picture of the sufficient sample complexity, let us study a simple example. Consider the simple case of a binary symmetric source $\bfX \stackrel{i.i.d.}{\sim} \text{Bern}(0.5)$ and binary symmetric channel with crossover probability $\delta$. That is, we consider black and white images, corrupted by a simple channel that flips pixels at random. For simplicity let us presume that $\gamma_n = n^{-1/2}$ and let $M = n^5$.

The minimum sample size, $N_{BSC}(\epsilon,\delta)$, that satisfies \eqref{eqn:achievability_condn} as a function of the error probability $\epsilon$ and the crossover probability $\delta$ is shown in Fig.~\ref{fig:second_order_bsc}. As expected, with decreasing target error probability, the sufficient sample size increases. Similarly as the channel gets noisier, that is, as the crossover probability increases, the sufficient sample size increases.

Thus, in this section we studied the channel-aware image registration problem in the finite-sample regime. The Feinstein decoder helps establish strong achievability conditions for the problem using finite-resolution images. Given the closeness of the Feinstein decoder to both the ML and MMI decoders, we argue that the result not only highlights the possible performance of universal methods like MMI, but also is fundamentally close to that achieved by the Bayes optimal ML decoder. In the next section we focus on bridging the gap between the universal and channel-aware contexts by deriving strong bounds on the performance of the MMI decoder, in terms of that of the ML decoder.

\section{Universal Delay Estimation: Relationship to Cycles} \label{sec:univ_delay_est}

The universal version of the image registration problem is complicated by two aspects---the growing number of hypotheses with sample size $n$, and the complicated transformations (permutations) applied to the images. In particular, properties such as the number of possible permutation cycles and fixed points affect the efficacy of the MMI decoder. Thus it is important to study the performance tradeoff as a function of these properties in the finite-sample context. We specifically study the tradeoff as a function of the maximum number of permutation cycles through the one-dimensional version of the problem, universal delay estimation \cite{SteinZM1996}. Here we perform a stricter analysis of the type counting argument to better emphasize the effect of the number of permutation cycles.

\begin{figure}[t]
	\centering
	\includegraphics[scale=0.75]{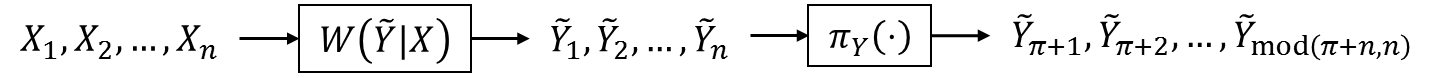}
 	\caption{Model of universal delay estimation: Source $\bfX$ is corrupted by channel $W$ and cyclically shifted by $\pi$ to obtain sequence $\bfY$.}
  	\label{fig:univ_delay_est}
\end{figure}

Specifically, we consider an $n$-length sequence of i.i.d.\ symbols $\bfX$ transmitted through an unknown channel $W$, and subject to a cyclic delay $\pi$ to obtain the received sequence $\bfY$, as shown in Fig~\ref{fig:univ_delay_est}. The goal here is to estimate the delay $\pi$. It is directly evident that this is a simplified version of the image registration problem, with one dimensional sources and a more limited set of possible transformations.

We now characterize the gap in performance of the MMI and ML decoders using a type counting argument. To do that, let us first define the delay types and some of their properties.

\subsection{Whittle's Law and Delay Types}

We now summarize a few results on the number of types and Markov types which are eventually used to define delay types. This is used to analyze the performance of the image registration algorithms. We follow the notation of \cite{Csiszar1998} and mention a few important results for completeness of the description.

Consider a sequence $\bfx \in \calX^n$. The empirical distribution $q_X$ of $\bfx$ is the \emph{type} of the sequence. Let $X \sim q_X$ be a dummy random variable used to reference the type. Let $T_X^n$ be the set of all sequences of length $n$, of type $q_X$. The number of possible types of sequences of length $n$ is polynomial in $n$, i.e., $O(n^{|\calX|})$ \cite{Csiszar1998}. The number of sequences of length $n$, of type $q_X$, is bounded as \cite{Csiszar1998}
\begin{equation} \label{eqn:type0_bound}
(n+1)^{-|\calX|}2^{nH(X)} \leq |T_X^n| \leq 2^{nH(X)}.
\end{equation}

The first-order Markov type of a sequence $\bfx \in \calX^n$ is defined as the empirical distribution $q_{X_0,X_1}$, given by
\[
q_{X_0,X_1}(a_0,a_1) = \frac{1}{n} \sum_{i=1}^n \indc{(x_i,x_{i+1}) = (a_0,a_1)}.
\]
Here we assume that the sequence is cyclic with period $n$, i.e., for any $i > 0, x_{n+i} = x_{i}$. Let $(X_0,X_1) \sim q_{X_0,X_1}$. The set of sequences of length $n$ with first order Markov type $q_{X_0,X_1}$ is given by the set $T_{X_0,X_1}^n$. From Whittle's theorem \cite{Whittle1955}, the size of the first order Markov type $T_{X_0,X_1}^n$ is bounded as
\begin{align}
|\calX|(n+1)^{-(|\calX|^2+|\calX|)} 2^{n(H(X_0,X_1) - H(X_0))} \leq |T_{X_0,X_1}^n| \leq |\calX| 2^{n(H(X_0,X_1) - H(X_0))}. \label{eqn:type1_bound}
\end{align}

The joint first-order Markov type of a pair of sequences $\bfx \in \calX^n$, $\bfy \in \calY^n$ is the empirical distribution
\[
q_{X_0,X_1,Y}(a_0,a_1,b) = \frac{1}{n} \sum_{i=1}^n \indc{(x_i,x_{i+1},y_i) = (a_0,a_1,b)}.
\]
Then given $\bfx$, the set of conditional first-order Markov type sequences, $T_{Y \vert X_0,X_1}^n(\bfx)$ are the sequences $\bfy$ such that $(\bfx,\bfy)$ has joint type $q$ where $(X_0,X_1,Y) \sim q$. This satisfies \cite{SteinZM1996}
\begin{align} 
(n+1)^{-|\calX|^2|\calY|} 2^{n(H(X_0,X_1,Y) - H(X_0,X_1))} \leq |T_{Y \vert X_0,X_1}^n(\bfx)| \leq 2^{n(H(X_0,X_1,Y) - H(X_0,X_1))}. \label{eqn:cond_Markov_type}
\end{align}

We generalize the notion of the Markov type to any delay $\pi$. We refer to these types as delay types, defined as follows.
\begin{defn}
For any delay $\pi$, and sequences $\bfx,\bfy$, the \emph{delay type}, $q_{X_0,X_{\pi}}$, and the \emph{joint delay type} $q_{X_0,X_{\pi},Y}$ are defined as 
\begin{align*}
q_{X_{0},X_{\pi}}(a_0,a_1) &= \frac{1}{n} \sum_{i=1}^n \indc{(x_i,x_{\text{mod}(\pi+i,n)}) = (a_0,a_1)}, \\
q_{X_{0},X_{\pi},Y}(a_0,a_1,b) &= \frac{1}{n} \sum_{i=1}^n \indc{(x_i,x_{\text{mod}(\pi+i,n)},y_i) = (a_0,a_1,b)}.
\end{align*}
\end{defn}

Consider $X_0,X_{\pi} \sim p$. Then the set of sequences $\bfx$ of length $n$ with delay type $p$ is given by  $T_{X_0,X_{\pi}}^n$. Given a sequence $\bfx$, the conditional type set $T_{Y\vert X_0,X_{\pi}}^n(\bfx)$ is the set of sequences $\bfy$ such that the joint delay type of $(\bfx,\bfy)$ is $q$ where $(X_0,X_{\pi},Y) \sim q$.  

We now characterize the size of the delay types. Let $\kappa_{\pi}$ be the number of cycles created by the shift. Then the length of each cycle is $\tfrac{n}{\kappa_{\pi}}$. Here we limit the delays just as in the image registration problem, such that $\kappa_{\pi} = o\pth{\tfrac{n}{\log n}}$. Then the size of the joint delay type is bounded as shown below.

\begin{lemma} \label{lemma:joint_type_bd_cyc}
For cyclic shift $\pi$, and sequence $\bfx$, we have
\begin{align}
\left| \log_2 |T_{X_0,X_{\pi}}^n| - n \pth{H(X_0,X_{\pi}) - H(X)} - \kappa_{\pi} \log r \right|  \leq \kappa_{\pi} r^2 \log_2 \pth{1+\frac{n}{\kappa_{\pi}}}, \notag
\end{align}
where $\kappa_{\pi}$ is the number of permutation cycles created by the cyclic shift $\pi$.
\end{lemma}
\begin{IEEEproof}
Recall the bounds of Whittle's theorem from \eqref{eqn:type1_bound}. Then, similar to \cite[Lem.~2]{RamanV2018c}, we derive the upper bound by studying the first-order Markov types over the cycles of the delay type. Let $q$ be the joint permutation type of $\bfx$. Let the first-order Markov type over cycle $i$ of the cyclic shift be $q_i$. Then, 
\begin{equation} \label{eqn:joint_type_cycle_decomp}
q(a_0,a_1) = \frac{1}{\kappa_{\pi}} \sum_{i=1}^{\kappa_{\pi}} q_i(a_0,a_1).
\end{equation}
Let $q',q_i'$ be the marginals corresponding to the joint types $q,q_i$ respectively.

Then, the size of the type class is obtained by summing over all valid decompositions in \eqref{eqn:joint_type_cycle_decomp} as
\begin{align}
|T_{ X_0,X_{\pi} }^n|  &= \sum \prod_{i=1}^{\kappa_{\pi} } |T_{q_i}^{n/\kappa_{\pi}}|  \label{eqn:perm_type_prod_Markov} \\
&\quad \leq \prod_{i=1}^{ \kappa_{\pi} } r \pth{ 1+\frac{n}{\kappa_{\pi}} }^{r^2} 2^{\frac{n}{\kappa_{\pi}} \pth{H(q_i) - H(q_i')}} \label{eqn:Whittle_ub} \\
&\quad \leq 2^{\qth{ n\pth{ \frac{1}{\kappa_{\pi}} \sum_{i=1}^{\kappa_{\pi}} (H(q_i) - H(q_i'))} + \kappa_{\pi} \pth{\log r + r^2 \log_2\pth{1+\tfrac{n}{\kappa_{\pi}} } } } }, \label{eqn:sum_bd_ub}
\end{align}
where \eqref{eqn:perm_type_prod_Markov} follows from the fact that the joint type is composed of a sequence from each of the first-order Markov types that compose the joint type in \eqref{eqn:joint_type_cycle_decomp}. Then, \eqref{eqn:Whittle_ub} follows from the upper bound on the size of the first-order Markov types by Whittle's law and from the fact that the number of types is polynomial. Finally, we bound the average of the entropy in \eqref{eqn:sum_bd_ub} as in \cite[Lem.~2]{RamanV2018c} to obtain the upper bound. 

The lower bound is obtained by the observation that the number of sequences from \eqref{eqn:perm_type_prod_Markov} is lower bounded by any one viable decomposition in \eqref{eqn:joint_type_cycle_decomp}. Thus, the lower bound follows the proof of \cite[Lem.~2]{RamanV2018c} \emph{mutatis mutandis}.
\end{IEEEproof}

Note that the result strengthens the conclusion of \cite[Lem.~2]{RamanV2018c}, making the dependence on the number of cycles more explicit. Similar results can be obtained for the conditional joint types as well, as shown below.
\begin{lemma} \label{lemma:cond_type_bd}
For any delay $\pi$, and any $\bfx,\bfy$, we have
\begin{align}
0 &\leq \log_2 |T_{Y \vert X_0,X_{\pi}}^n| - n \pth{H(X_0,X_{\pi},Y) - H(X_0,X_{\pi})} \leq \kappa_{\pi} r^3 \log_2 \pth{1+\frac{n}{\kappa_{\pi}}} .
\end{align}
\end{lemma}
\begin{IEEEproof}
The proof is similar to that of Lemma \ref{lemma:joint_type_bd_cyc}.
\end{IEEEproof}

We use the stronger bounds on the sizes of the delay type sets to derive performance bounds of the image registration algorithms next.

\subsection{Performance Analysis}

The error probability of the MMI decoder can now be bounded more precisely as follows.
\begin{theorem} \label{thm:MMI_strict_bounds}
Let the maximum number of permutation cycles, given the set of possible delays, be $\kappa$. Then, 
\begin{align}
-\frac{1}{n}\log_2 P_e(\Phi_{\text{MMI}}) &\geq \calE^* - \frac{\kappa}{n} \pth{r^2(r+1)} \log_2\pth{1+\tfrac{n}{\kappa}} - r^3\frac{\log_2(1+n)}{n}  - \frac{\kappa}{n}\log r - \frac{1}{n}\log \kappa,
\end{align}
where $\calE^*$ is the error exponent.
\end{theorem}
\begin{IEEEproof}
The result for the error in the binary hypothesis test with respect to the null hypothesis and a delay of $\pi$ is analogous to that of \cite[Thm.~2]{RamanV2018c}. The result is obtained by substituting the tighter bounds derived in Lemmas~\ref{lemma:joint_type_bd_cyc} and \ref{lemma:cond_type_bd}, and the observation that the number of joint types over $X_0,X_{\pi},Y$ are bounded by $(n+1)^{r^3}$. Finally, using union bound results in the bound.
\end{IEEEproof}

Similarly we can obtain the converse by studying the performance of the maximum likelihood decoder.
\begin{theorem} \label{thm:ML_strict_bound}
Let the maximum number of permutation cycles, given the set of possible delays, be $\kappa$. Then, 
\begin{align}
&-\frac{1}{n}\log_2 P_e(\Phi_{\text{ML}}) \leq \calE^* + \frac{\kappa}{n} \pth{r^2} \log_2\pth{1+\tfrac{n}{\kappa}}  - \frac{\kappa}{n}\log r.
\end{align}
\end{theorem}
\begin{IEEEproof}
The proof is analogous to that of Thm.~\ref{thm:MMI_strict_bounds} and follows from Lems.~\ref{lemma:joint_type_bd_cyc} and \ref{lemma:cond_type_bd}.
\end{IEEEproof}

Theorems \ref{thm:MMI_strict_bounds} and \ref{thm:ML_strict_bound} make explicit the performance tradeoff as a function of the number of permutation cycles. The suboptimality of MMI in terms of ML is better characterized as a function of the number of permutation cycles $\kappa$ as shown below.

\begin{corollary}
The suboptimality of MMI is characterized in terms of the performance of ML as
\begin{align}
0 &\leq \frac{1}{n} \qth{\log_2 P_e(\Phi_{\text{MMI}}) - \log_2 P_e(\Phi_{\text{ML}}) } \notag \\
&\quad \leq \frac{\kappa}{n}(r^2(r+2)) \log_2\pth{1+\frac{n}{\kappa}} + r^3 \frac{\log_2(1+n)}{n} + \frac{1}{n} \log_2 \kappa.
\end{align}
\end{corollary}
\begin{IEEEproof}
This is a direct consequence of Thms.~\ref{thm:MMI_strict_bounds} and \ref{thm:ML_strict_bound}.
\end{IEEEproof}

We observe that the dominant higher-order term is $O(\kappa\log_2 n)$ reinforcing the fact that for $\kappa = o\pth{\frac{n}{\log n}}$, the exponents match that of the maximum likelihood decoder \cite{RamanV2018c}. 

The larger the number of permutation cycles, the higher the upper bound in the difference of performance of MMI and ML. To understand why, we look at the functioning of the MMI decoder. The MMI method essentially performs an independence testing of the samples across the set of possible transformations. The samples across different cycles are independent in the computation anyway. 

Further, the larger the value of $\kappa$, the smaller each cycle. Thus, the information computation translates to computing the information on $\kappa$ i.i.d.\ sets of size $n/\kappa$ each. Thus, when $\kappa$ is large, the information estimates on each cycle is less accurate, therein making the estimator less reliable as well. This effect is particularly more prominent in the universal context as the ML decoder is aware of the channel, and thus more robust.

The result characterizes the loss in performance from the lack of knowledge of the statistics of the channel, i.e., a bound on the cost of universality. Through a stricter analysis of the type counting argument in order to characterize the effect of the number and size of permutation cycles on the performance of the MMI method in universal delay estimation. Whereas the finite-sample analysis for universal methods proves to be hard, such tighter type counting arguments provide stronger insight into the relationship between the nature of the transformations and algorithm performance.

\section{Conclusion} \label{sec:conclusion}

In this paper we studied the image registration problem with a focus on tighter performance analyses in the finite-sample context. First, we considered the channel-aware version of the problem wherein we defined the Feinstein decoder for image registration and computed achievable error rates in the finite sample setting using the Berry-Esseen CLT. Future work can build on strong large deviations and central limit theorems to obtain converse criteria for image registration.

Then, we studied the universal delay estimation to characterize the higher-order terms in the difference in performance of the MMI method from that of ML. This gave us a stronger characterization of the tradeoff as a function of the properties of the transformations. This sheds light on the finite-sample performance in the universal setting by benchmarking them against the channel-aware context.

In the future we hope to build on the current work by developing a strong converse argument that leverages strong large deviations \cite{Moulin2017} to compute the finite sample lower bound on the error probability. Acccordingly strengthening the achievability result could help establish the fundamental limits of channel-aware image registration in the finite sample context. Another direction to explore further is the finite-sample analysis of universal image registration algorithms such as MMI. One possible approach to explore is to consider the finite-sample performance of functional estimates of probabilities \cite{IngberWK2012}. Alternately, one could also benchmark the performance of MMI against that of the Feinstein decoder using \eqref{eqn:KL_Diff_density} by establishing strong finite sample bounds on the KL divergence between the empirical and true distributions.

Beside theoretical analyses, the effectiveness of these algorithms like the MMI and the Feinstein decoder also inspire the definition of efficient practical heuristics that leverage the variational characterization of information functionals and the empirical gradient estimates. The efficacy of deep neural models and the definition of neural network based information estimators \cite{BelghaziBROBCH2018} further expand the horizon for us to explore neural implementations of MMI-type image registration algorithms that estimate the image transformations in the continuous space efficiently.

Such information theoretic explorations of classical statistical inference problems not only help define the fundamental limits and establish the benchmarks for practical algorithms to strive for, but also inspire the definition of efficient and novel algorithms. 

\bibliographystyle{IEEEtran}
\bibliography{abrv,conf_abrv,lrv_lib}

\end{document}